\documentclass[12pt]{article}

\begin{document}
\begin{titlepage}
\begin{center}

{\large \bf {LEPTONIC CP VIOLATION and NEUTRINO MASS MODELS}}

\vskip 1cm

Gustavo C. Branco\footnote{gbranco@gtae3.ist.utl.pt} 
and
M. N. Rebelo  
\footnote{rebelo@alfa.ist.utl.pt}
\vskip 0.05in

{\em Departamento de F{\'\i}sica and Centro  de F{\'\i}sica
Te{\'o}rica de Part{\'\i}culas (CFTP),\\
Instituto Superior T\'{e}cnico, Av. Rovisco Pais, 1049-001
Lisboa, Portugal}

\end{center}

\vskip 3cm

\begin{abstract}
We discuss leptonic mixing and CP violation at low and
high energies, emphasizing possible connections between
leptogenesis and CP violation at low energies, in the
context of lepton flavour models. Furthermore we analyse weak basis
invariants relevant for leptogenesis and for CP violation 
at low energies. These invariants have the advantage of
providing a simple test of the CP properties of any lepton
flavour model.
\end{abstract}

\end{titlepage}

\newpage
\section{Introduction}
The experimental data on atmospheric and solar neutrinos have provided
evidence for non-vanishing neutrino masses and for non-trivial leptonic
mixing \cite{parametros}. 
These important discoveries rendered even more pressing
the fundamental question 
of understanding the spectrum of fermion masses and the 
pattern of their mixing. 
In the Standard Model (SM) neutrinos are strictly massless.
No Dirac masss terms can arise in the SM due to the absence of 
right-handed (rh) neutrinos and no left-handed (lh) Majorana masses can
be generated at tree level due to the simple Higgs structure of the SM.
Furthermore, no Majorana masses can be generated in higher orders
due to the exact B-L conservation. Therefore, the discovery of neutrino
masses and leptonic mixing provides clear evidence for Physics beyond 
the SM. 

It is remarkable that a simple extension of the SM, through 
the introduction of rh neutrinos, leads to non-vanishing but naturally
small neutrino masses. With the addition of rh neutrinos to the SM,
the most general Lagrangean consistent with renormalizability and gauge
invariance leads to both Dirac and rh Majorana neutrino 
mass terms. The natural scale for the Dirac neutrino masses is v, the
scale of electroweak symmetry breaking. On the other hand, since the
rh neutrinos 
transform trivially under SU(2)x U(1), the rh Majorana mass term 
is gauge invariant and as a result its scale V can be much larger,
being
identified with the scale of lepton number violation. In the context of
Grand Unified  Theories (GUT) this scale can be naturally taken as the 
GUT scale. The presence of both Majorana and Dirac masses of the above
indicated order of magnitude, automatically leads to light neutrinos
with masses of order $v^2/V$, through the seesaw 
mechanism \cite{seesaw}. Strictly speaking,
in order to have naturally small neutrino masses  
it is not necessary to introduce rh neutrinos, one may have only lh
neutrinos, provided lepton number violation occurs at a high energy
scale.
The introduction of rh neutrinos is well motivated in the framework
of some GUT theories like S0(10) and it has the special 
appeal of establishing a 
possible connection between neutrinos and the generation of the baryon
asymmetry of the universe (BAU). In fact, one of the most attractive
mechanisms to generate BAU is baryogenesis through 
leptogenesis \cite{Fukugita:1986hr}, a
scenario where the out of equilibrium decays of heavy rh neutrinos 
create a lepton asymmetry which is later converted into a baryon
asymmetry by B+L violating (but B-L conserving) sphaleron
interactions \cite{Kuzmin:1985mm}.    
    
   It is well known \cite{Grimus:1995zi} that pure gauge theories 
do not violate CP. In fact, the fermionic sector (kinetic energy 
terms and fermion interactions with vector bosons) as well as the 
vector boson sector of gauge theories are always CP symmetric. The 
same is true for the couplings of scalars with gauge fields. 
In the SM, CP violation in the quark sector arises from the 
simultaneous presence of charged current gauge interactions 
and complex Yukawa couplings \cite{Branco:1999fs}. 
In general, for three or more generations there is no CP transformation
which leaves invariant both the Yukawa couplings and the charged 
current gauge interactions. This leads to the well known 
Kobayashi-Maskawa mechanism   
of CP violation operating in the quark sector. In the leptonic sector 
and in the context of  
the SM, there is no CP violation since for massless neutrinos leptonic
mixing in the charged currents can always be rotated away through a
redefinition of neutrino fields. In any extension of the SM with
non-vanishing neutrino masses and mixing, there is in general 
leptonic CP violation. In the case of an extension of the SM 
consisting of the addition three rh neutrinos, one has in general both 
leptonic CP violation at low energies, visible for example through 
neutrino oscillations and CP violation at high energies relevant for the
generation of baryogenesis through leptogenesis.

   In this paper we review leptonic mixing and CP violation at low
and high energies, with emphasis on the possible connection between
leptogenesis and low energy data as well as on the analysis 
of weak-basis (WB) invariants relevant for CP violation. 
In fact by writing the most general CP transformation  
for the fermion fields in a weak basis 
one can derive simple conditions for CP conservation
which can be applied without going to the physical basis. This
strategy was followed for the first time in the context of the
Standard Model in Ref. \cite{Bernabeu:1986fc}. These
invariants provide a simple way of testing whether a specific
lepton flavour model \cite{Altarelli:2002hx}
leads to CP violation either at low or
high energies. The crucial advantage of these invariants
stems from the fact that for any lepton flavour model, they can be 
calculated in any WB, without requiring cumbersome changes of basis.
The paper is organized as follows.
In section 2 we establish our notation introducing the various leptonic 
mass terms, derive necessary conditions for CP invariance and 
identify the independent CP violating phases, both in a 
WB and in the mass eigenstate basis. In section 3, we derive WB
invariants which are relevant for CP violation at low energies,
as well as WB invariants sensitive to CP violation at high energies
relevant for leptogenesis. In section 4, we analyse the special
limit of exactly degenerate neutrino masses. The relationship
between low energy CP violation and CP violation at high 
energies is discussed in section 5. Finally, in section 
6, we present our summary and conclusions.

\section{Neutrino Mass Terms}
\setcounter{equation}{0}
We consider a simple extension of the SM where three rh neutrinos
(one per generation) are introduced. In this case, the most
general form for the leptonic mass terms after spontaneous
symmetry breaking is:
\begin{eqnarray}
{\cal L}_m  &=& -[\frac{1}{2} \nu_{L}^{0T} C m_L \nu_{L}^0+
\overline{{\nu}_{L}^0} m_D \nu_{R}^0 +
\frac{1}{2} \nu_{R}^{0T} C M_R \nu_{R}^0+
\overline{l_L^0} m_l l_R^0] + h. c. = \nonumber \\
&=& - [\frac{1}{2}  n_{L}^{T} C {\cal M}^* n_L +
\overline{l_L^0} m_l l_R^0 ] + h. c.
\label{lep}
\end{eqnarray}
where $m_L$, $M_R$, denote the lh
and rh neutrino Majorana mass matrices, 
while $m_D$, $m_l$ stand for 
the neutrino Dirac mass matrix and the charged
lepton mass matrix, respectively. 
The generation at tree level of a mass term 
of the form $\nu_{L}^{0T} C m_L \nu_{L}^0$
also requires the extension of the Higgs sector (e.g., a Higgs triplet).
The introduction
of the column vector $n_L = ({\nu}_{L}^0, {(\nu_R^0)}^c)$ 
allows one to write ${\cal L}_m $ in a more compact form, with
the $6 \times 6$ matrix $\cal M $  given by:
\begin{eqnarray}
{\cal M}= \left(\begin{array}{cc}  
m^*_L  & m_D \\
m^T_D & M_R \end{array}\right) \label{calm}
\end{eqnarray}
 
The mass terms in  ${\cal L}_m$ contain all the 
information on CP violation 
arising from the charged gauge interactions, irrespective 
of the mechanism which generates the lepton mass terms
and will be analysed in the next subsection. An enlarged
Higgs sector will in general provide new sources of CP
violation which we do not discuss in this work.
In fact most of our analysis will be done
in the framework of the minimal Higgs structure
(no Higgs triplets), thus implying that the term in
$m_L $ in Eq. (\ref{lep}) is absent.
The corresponding matrix $\cal M $ has then a zero block 
entry in its upper left block.

For simplicity, in most of this paper we will consider
that the number of rh neutrinos equals the number of 
lh neutrinos. It should be pointed out that this is not 
required in order for appropriate neutrino masses to be
generated.

\subsection{The general case}
In this subsection we study leptonic CP violation in the case
corresponding to the most general mass terms given by 
Eq. (\ref{lep}). There are two aspects in which leptonic
CP non-conservation differs from CP violation in the
quark sector, One aspect has to do with the fact that
being neutral, neutrinos can have both Majorana and Dirac 
mass terms. The other one results from the fact that
the full leptonic mixing matrix appearing in the charged currents
is a $3 \times 6 $ matrix, consisting of the first three
lines of a $6 \times 6$ unitary matrix. Of course, in the 
low energy limit, where only the light neutrinos are active,
the leptonic mixing is described by a $3 \times 3$ unitary matrix. 
For the analysis of leptonic mixing and CP violation 
mediated through the charged gauge bosons
the relevant part of the Lagrangean is
${\cal L}_m$ given by Eq. (\ref{lep}) together with the
charged gauge interaction
\begin{equation}
{\cal L}_W = - \frac{g}{\sqrt{2}}  W^+_{\mu} \ 
 \overline{l^0_L} \ \gamma^{\mu} \ \nu^0_{L} +h.c.
\label{16}
\end{equation}
The simplest  way of determining the number of independent
CP violating phases  \cite{Branco:1986gr}
is by working in a conveniently chosen weak basis
(WB) and analysing the restrictions on the Lagrangean implied by
CP invariance. We follow this approach, but also identify
the CP violating phases appearing in the charged weak interactions, 
written in the mass eigenstate basis.

The most general CP transformation which leaves the gauge interaction
invariant is:
\begin{eqnarray}
{\rm CP} l^0_L ({\rm CP})^{\dagger}&=&U^\prime
\gamma^0  C~ \overline{l^0_L}^T;  \quad
{\rm CP} l^0_R({\rm CP})^{\dagger}=V^\prime \gamma^0  C~ \overline{l^0_R}^T
\nonumber \\
{\rm CP} \nu^0_L ({\rm CP})^{\dagger}&=&U^\prime \gamma^0 C~
\overline{\nu^0_L}^T; \quad
{\rm CP} \nu^0_R ({\rm CP})^{\dagger}=W^\prime \gamma^0  C~
\overline{\nu^0_R}^T \label{cp}  \\
{\rm CP} W^+_{\mu} ({\rm CP})^{\dagger} &=& - (-1)^{\delta_{0\mu}}  
W^-_{\mu} \nonumber
\end{eqnarray}
where $U^\prime$, $V^\prime$, $W^\prime$ are unitary matrices 
acting in flavour space. This transformation combines the CP
transformation of a single fermion field with a WB transformation.
Invariance of the mass terms under the above CP transformation,
requires that the following relations have to be satisfied:
\begin{eqnarray}
U^{\prime T} m_L U^\prime &=&-m_L^*  \label{cpll} \\
W^{\prime T} M_R W^\prime &=&-M_R^*  \label{cpM} \\
U^{\prime \dagger} m_D W^\prime &=& {m_D}^*  \label{cpm} \\
U^{\prime \dagger} m_l V^\prime &=& {m_l}^* \label{cpml}
\end{eqnarray}
It can be easily seen that if there are unitary matrices 
$U^\prime$, $V^\prime$, $W^\prime$ satisfying 
Eqs. (\ref{cpll}) -- (\ref{cpml}) in one particular WB, then 
a solution exists for any other WB.
In order to analyze the implications of the above conditions,
it is convenient to choose the WB where both $m_L$ and $M_R$ are real
diagonal. In this WB and assuming the eigenvalues of $m_L$ and $M_R$ 
to be all non-zero and non-degenerate, 
Eqs. (\ref{cpll}) and (\ref{cpM}) constrain  $U^\prime$ and
$W^\prime$ to be of the form:
\begin{eqnarray}
U^\prime={\rm diag.} \left(\exp(i\alpha_1), 
\exp(i\alpha_2),... \exp(i\alpha_n)\right)  \label{expu}\\
W^\prime ={\rm diag.} \left(\exp(i\beta_1), 
\exp(i\beta_2),... \exp(i\beta_n)
\right) \label{expw}
\end{eqnarray}
where n denotes the number of generations. Here we are assuming, 
for simplicity, that there is an equal number of fields $\nu^0_{L}$
and $\nu^0_{R}$. The phases $\alpha_i$  and $\beta_i$ have to
satisfy:
\begin{equation}
\alpha_i=(2 p_i +1) \frac{\pi}{2}, \quad
\beta_i=(2 q_i +1) \frac{\pi}{2} \label{phs}
\end{equation}
with $p_i$, $q_i$  integer numbers. Then Eqs. (\ref{cpm}) and 
(\ref{cpml}) constrain $m_D$ and $m_lm_l^\dagger \equiv h_l$ in the
following way:
\begin{eqnarray}
{\rm phase} (m_D)_{ij} = (p_i - q_j)  \frac{\pi}{2} \\
{\rm phase} (h_l)_{ij} = (p_i - q_j)  \frac{\pi}{2} 
\label{way}
\end{eqnarray}
As a result, CP invariance restricts all the phases of 
$m_D$ and  $h_l$ to be either zero or $\pm \pi /2$. 
Since in general $m_D$ is an arbitrary $n \times n$ complex matrix
whilst $h_l$ is an arbitrary  $n \times n$ Hermitian matrix the
number of independent CP restrictions is:
\begin{equation}
N_g = n^2 + \frac{n(n-1)}{2}
\end{equation}
For three generations $N_g = 12$. It is clear that if the number of
righthanded fields where $n^\prime$ rather than n the matrix
$m_D$ would have dimension  $n \times n^\prime $ and $N_g$
would be given by 
\begin{equation}
N^{\prime}_g = nn^\prime + \frac{n(n-1)}{2}.
\label{nova}
\end{equation}

It can be checked that this number of CP restrictions
coincides with the number of CP violating phases which
arise in the leptonic mixing matrix of the charged weak
current after all leptonic masses have been diagonalized.
Let us now choose the WB where $m_l$ is already diagonal, real 
and positive. The diagonalization of the $2n \times 2n$
matrix $\cal M$, which in general is given by Eq. (\ref{calm}), 
is performed via the unitary transformation
\begin{equation}
V^T {\cal M}^* V = \cal D \label{dgm}
\end{equation}
where ${\cal D} ={\rm diag.} (m_{\nu_1}, m_{\nu_2}, m_{\nu_3},
M_{\nu_1}, M_{\nu_2}, M_{\nu_3})$,
with $m_{\nu_i}$ and $M_{\nu_i}$ denoting the physical
masses of the light and heavy Majorana neutrinos, respectively. It is
convenient to write $V$ and $\cal D$ in the following block form:
\begin{eqnarray}
V&=&\left(\begin{array}{cc}
K & R \\
S & T \end{array}\right) ; \label{matv}\\
{\cal D}&=&\left(\begin{array}{cc}
d & 0 \\
0 & D \end{array}\right) . \label{matd}
\end{eqnarray}
The neutrino weak-eigenstates are
related
to the mass eigenstates by:
\begin{eqnarray}
{\nu^0_i}_L= V_{i \alpha} {\nu_{\alpha}}_L=(K, R)
\left(\begin{array}{c}
{\nu_i}_L  \\
{N_i}_L \end{array} \right) \quad \left(\begin{array}{c} i=1,2,3 \\
\alpha=1,2,...6 \end{array} \right)
\label{15}
\end{eqnarray}
and thus the leptonic charged current interactions are given by:
\begin{equation}
{\cal L}_W = - \frac{g}{\sqrt{2}} \left( \overline{l_{iL}} 
\gamma_{\mu} K_{ij} {\nu_j}_L +
\overline{l_{iL}} \gamma_{\mu} R_{ij} {N_j}_L \right) W^{\mu}+h.c.
\label{phys}
\end{equation}
with $K$ and $R$ being the charged current couplings of charged 
leptons to the light neutrinos $\nu_j$ and to the heavy neutrinos 
$N_j$, respectively. From Eq. (\ref{matv}) we see that $K$ and $R$ 
correspond to the first n rows of the $2n \times 2n$ unitary
matrix V which diagonalizes the full neutrino mass matrix
$\cal M^*$. The most general $n \times 2n$ leptonic mixing 
matrix can then be exactly parametrized by the first n rows of a 
$2n \times 2n$ unitary matrix provided that it is chosen in such a way
that a minimal number of phases appears in these first $n$ rows.
This is the case of the parametrization proposed in 
Ref \cite{Anselm:1985rw}. Its particularization for a 
$6 \times 6$ matrix is given by:
\begin{eqnarray}
V= {\hat V} P  \label{vip}
\end{eqnarray}
where $P={\rm diag.}
\left(1, \exp(i\sigma_1), \exp(i\sigma_2),..., \exp(i\sigma_5)
\right) $ and  $\hat V$ is given by:
\begin{eqnarray}
{\hat V}= & O_{56}I_6(\delta_{10}) O_{45} O_{46} I_5 (\delta_9)
I_6(\delta_8) \left ( \prod_{j=4}^6  O_{3j}\right )  I_4 (\delta_7) 
I_5 (\delta_6) I_6 (\delta_5 ) \times \nonumber \\
 & \times \left ( \prod_{j=3}^6 O_{2j} \right ) I_3 (\delta_4) I_4 (\delta_3)
I_5(\delta_2) I_6(\delta_1)  \left ( \prod_{j=2}^6
 O_{1j} \right )  \label{vcp}
\end{eqnarray}
where $O_{ij}$ are orthogonal matrices mixing
the ith and jth generation
and $I_j(\delta_k)$ are unitary diagonal matrices of the form:
\begin{eqnarray}
I_j(\delta_k)=\left(\begin{array}{cccccccc} \\
1&&&&&& \\
&.&&&&& \\
&&1&&&& \\
&&&e^{i \delta_k}&&&\\
&&&&1&& \\
&&&&&.& \\
&&&&&&1 \\
\end{array} \right) \leftarrow j
\label{ij}
\end{eqnarray}
This parametrization is particularly useful, for
instance, in models with vectorial quarks \cite{Branco:1986my}.
It can be readily verified that the first three rows of
$\hat V$, contain seven phases. The Majorana character of the
physical neutrinos does not allow for the five phases in P to
be rotated away and we are finally left with twelve phases
in the mixing matrix $(K,R)$. The generalization to $n + n^\prime$
dimensional unitary matrices leads to 
$\frac{1}{2} (n-1)(n-2+2n^\prime )$ phases in the first
n rows of $\hat V$ \cite{Branco:1986my} which, together with the 
$(n + n^\prime - 1)$ phases that cannot be rotated away,
adds up to  $nn^\prime + \frac{n(n-1)}{2}$ thus coinciding with 
the general result obtained in Eq. (\ref{nova}).

\subsection{The case of minimal seesaw}
The minimal seesaw case corresponds to ${\cal L}_m$ with no 
left-handed Majorana mass terms  
included, together with the assumption that the bare 
right-handed Majorana 
mass terms are much larger than the weak scale.
From Eqs. (\ref{calm}), (\ref{dgm}),  (\ref{matv}) and (\ref{matd}), 
with $m_L = 0$, one obtains:
\begin{eqnarray}
S^\dagger m^T_D K^* + K^\dagger m_D S^*
+ S^\dagger M_R S^* &=&d \label{12a} \\
S^\dagger m^T_D R^* + K^\dagger m_D T^*
+ S^\dagger M_R T^* &=&0 \label{12b} \\
T^\dagger m^T_D R^* + R^\dagger m_D T^*
+ T^\dagger M_R T^* &=&D  \label{12c}
\end{eqnarray}
We assume, as before, that we are already in a WB where $m_l$ is
real and diagonal. These equations allow us to derive the following
relations which hold to an excellent approximation:
\begin{eqnarray}
S^\dagger=-K^\dagger m_D M^{-1}_R \label{13} \\
-K^\dagger m_D \frac{1}{M_R} m^T_D K^* =d \label{14}
\end{eqnarray}
It is clear from 
Eq. (\ref{13}) that S is of order $m_D / M_R$ and therefore is 
very suppressed. Eq. (\ref{14}) is the usual seesaw formula
with the matrix $K$ frequently denoted by $V_{PMNS}$,
the Pontecorvo, Maki, Nakagawa, Sakata matrix \cite{pmns}.
Although the block $K$ in Eq. (\ref{matv}) is not a unitary matrix 
its deviations from unitarity are of the order $m^2_D / M^2_R$.
It is from  Eq. (\ref{14}) that the low energy physics of
the leptonic sector is derived. The decoupling limit
corresponds to an effective theory with only left-handed neutrinos
and a Majorana mass matrix, $m_{eff}$ defined as:
\begin{equation}
m_{eff} = - m_D \frac{1}{M_R} m^T_D \label{meff}
\end{equation} 
showing that for $m_D$ of the order of the electroweak scale and $M_R$
of the scale of grand unification, the smallness of light
neutrino masses is a natural consequence of the seesaw 
mechanism \cite{seesaw}.
From the relation ${\cal M}^* V = V^*  \cal D $ and taking into
account the zero entry in ${\cal M}$ one derives the following
exact relation
\begin{equation}
R=m_D T^* D^{-1} \label{exa}
\end{equation} 
This equation plays an important r\^ ole in the connection between
low energy and high energy physics in the leptonic sector. If we
choose to work in a WB where both $m_l$ and $M_R$ are diagonal,
Eq. (\ref{12c}) shows that $T= 1 $ up to corrections of order 
$m^2_D / M^2_R$, leading to an excellent approximation to
\begin{equation}
R=m_D D^{-1} \label{app}
\end{equation} 
The matrices $K$ and $R$ are again the charged current couplings.
The counting of the number of physical CP violating phases can 
be done in various ways  \cite{Korner:1992zk}, \cite{Endoh:2000hc}
\cite{Branco:2001pq}. The simplest approach \cite{Branco:2001pq}
is by choosing a WB where $M_R$ and  $m_l$
are simultaneously real and diagonal. From the
CP transformations given by Eq. (\ref{cp}) we now obtain
conditions of Eqs. (\ref{cpM}), (\ref{cpm}) and (\ref{cpml}). 
Once again, Eq. (\ref{cpM}) constrains the matrix $W^\prime$
to be of the form of Eq. (\ref{expw}) with $\beta _i$ given
by Eq. (\ref{phs}).  Multiplying Eq. (\ref{cpml}) by its Hermitian
conjugate, with $m_l$  real and diagonal, one concludes that 
$U^\prime $ has 
to be of the form of Eq. (\ref{expu}) where in this case
the  $\alpha_i$ are arbitrary phases. From Eqs. (\ref{cpm}),
(\ref{expw}), and (\ref{expu})  it follows then  
that CP invariance constrains the matrix $m_D$ to satisfy:
\begin{equation}
{\rm arg}(m_D)_{ij}=\frac{1}{2}(\alpha_i-\beta_j) \label{arg} 
\end{equation} 
Note that the $\beta_i$ are fixed up to discrete
ambiguities whilst the $\alpha_i$ are free. 
Therefore CP invariance constrains the matrix $m_D$
to have only $n$ free phases $\alpha_i$. Since $m_D$ is an arbitrary
matrix, with $n^2$ independent phases, it is clear that 
the number of independent CP restrictions is given by:
\begin{equation}
N_m = n^2-n
\label{min}
\end{equation}
In the minimal seesaw model,
for three generations, there are six CP violating phases 
instead of the twelve of the general case. The decrease in the 
number of independent phases is to be expected since in this case
$m_L$, which in general is a complex symmetric matrix and would have
six phases for three generations, is not present in the theory. 
We may still use the explicit parametrization given before by
Eqs. (\ref{vip}) and (\ref{vcp}). Yet, now the angles and phases
introduced are no longer independent parameters, there will be 
special constraints among them. The number of mixing angles
\cite{Endoh:2000hc} is also $(n^2-n)$, i.e., six mixing angles
for three generations. The exact form of these constraints 
can be derived from ${\cal M}^*  = V^*  {\cal D}  V^\dagger $
taking into account that  ${\cal M}$ has a zero entry
in the upper left block, which implies:
\begin{equation}
K^* d K^\dagger + R^* D R^\dagger =0. \label{krd}
\end{equation} 
An important physical question is how to distinguish experimentally
minimal seesaw from the general case. This is obviously a very
difficult (if not impossible) task, since it would require the
knowledge of the heavy neutrino masses as well as a detailed
knowledge of the matrix $R$. So far, we have not made any
assumption on the type of hierarchy in the light neutrino
masses (i.e. normal hierarchy, inverted hierarchy or almost degeneracy).
Recently it was argued that in grand unified 
models with minimal seesaw inverted hierarchy
for light neutino masses is theoretically 
disfavoured \cite{Albright:2004kb}. 

At this stage, it is useful to compare the 
number of physical parameters - three light and three heavy neutrino
masses, three charged lepton masses, six mixing angles and six
CP violating phases, giving a total of twenty one parameters - to the
number of parameters present 
in the WB where $M_R$ and $m_l$ are simultaneously real
and diagonal. In this case these two matrices contain six
real parameters. Since $m_D$ is a three by three general 
matrix, it contains nine real parameters and six phases due 
to the possibility of rotating away three phases on its 
left-hand side.Thus there are also twenty one parameters
in this WB. Obviously, not all  WB have the 
property of containing the minimum number of parameters. 
It is useful to parametrize  $m_D$ as a product of 
a unitary matrix U times a Hermitian matrix H, which can be 
done without loss of generality:
 \begin{equation}
m_D= UH =
P_{\xi}{\hat U_{\rho}} P_{\alpha} {\hat H}_{\sigma} P_{\beta}
\label{upy}
\end{equation}
In the second equality a maximum number of phases were factored
out of U and H leaving them with one phase each - $\rho $ and
$\sigma $ respectively, and  
$P_{\xi}={\rm diag.}(\exp(i\xi_1),$ $\exp(i\xi_2),$ $\exp(i\xi_3))$,   
$P_\alpha =diag.(1, \exp(i\alpha_1),$ $ \exp(i\alpha_2))$ 
and $P_\beta =diag.(1, \exp(i\beta_1),$ $ \exp(i\beta_2))$. The phases
in $P_{\xi}$ can be eliminated by rotating simultaneously
$\nu ^0_L$ and $l^0_L$. 
Alternatively one may write $m_D$,
without loss of generality, as the product of a unitary times 
a lower triangular matrix \cite{Hashida:1999wh}. This choice may 
be particularly useful in specific scenarios and it is 
easy to  show how the six independent phases may be chosen
\cite{Branco:2001pq}.

\section{WB invariants and CP violation} 
In this section we derive simple conditions for CP conservation
in the form of WB invariants which have to vanish in order for CP 
invariance to hold. These conditions are very useful, since they 
allow us to determine whether or not a given Lagrangean violates 
CP without the need to go to any special WB or to the physical basis.
This is specially relevant in the analysis of lepton flavour 
models, where the various matrices of Yukawa couplings may have
special textures in flavour space reflecting, for example,
the existence of a lepton flavour symmetry. In the presence of
texture zeros, WB invariants provide the simplest method
to investigate whether a specific lepton flavour model leads to
leptonic CP violation at low energies or whether the model
allows for CP violation at high energies, 
necessary to generate BAU through leptogenesis.
 
The method to build  WB invariants 
relevant for CP violation, was first proposed in
\cite{Bernabeu:1986fc} to the quark sector and was soon afterwards
extended to the low energy physics of the leptonic sector
\cite{Branco:1986gr}; the WB invariant relevant for
CP violation with three degenerate light
neutrinos was obtained  later in Ref \cite{Branco:1998bw}.
In reference \cite{Branco:2001pq} similar
conditions relevant for leptogenesis in the minimal seesaw
model with three generations were derived. This approach has
been widely applied in the literature \cite{many1}
to the study of CP violation in many different scenarios. 

It was shown in the previous section that CP invariance of
the charged gauge currents requires the existence of unitary
matrices  $U^\prime$, $V^\prime$, $W^\prime$ satisfying  
Eqs. (\ref{cpll}) -- (\ref{cpml}) or  just (\ref{cpM}) -- (\ref{cpml})
depending on whether $m_L$ is introduced. These matrices
have different forms in different WB. On the other hand, physically 
meaningful quantities must be invariant under WB transformations.
In order to derive conditions for CP invariance expressed in
terms of WB invariants we combine these equations 
in a non-trivial way and eliminate the dependence on the above
unitary matrices by using the fact that traces and determinants
are invariant under similarity transformations.
In the next subsections, we present and discuss conditions
relevant for different physical situations.

\subsection{WB Invariants relevant for CP Violation at Low Energies}
The different terms of ${\cal L}_m$ lead to conditions 
(\ref{cpll}) -- (\ref{cpml}) for CP invariance. The strategy
outlined above can be applied directly 
to this Lagrangean \cite{Branco:1986gr} leading 
among other interesting possibilities, to the 
following WB invariant CP conserving condition:
\begin{equation}
{\rm tr} \left[ (m^*_L m_L)^a, {h_l}^b \right]^q = 0
\label{trr}
\end{equation}
with $ h_l = m_l m^\dagger_l $, a, b, q integers and q odd. An analogous 
condition with $m_L$ and $h_l$ replaced by $M_R$ and 
$ h_D = m^\dagger_D m_D $ also holds. 
In the framework of minimal seesaw, $m_L$ is not present at
tree level. However, the low energy limit
of the minimal seesaw corresponds to an effective theory with only
left-handed neutrinos, with an effective  
Majorana mass matrix $m_{eff}$ given by
Eq. (\ref{meff}) in terms 
of $m_D$ and $M_R$. Invariance under CP of the
effective Lagrangean implies the following condition for
$m_{eff}$:
\begin{equation}
U^{\prime \dagger} m_{eff} U^{\prime *} = -m^*_{eff}  
\label{ueff}
\end{equation}
which is analogous to Eq. (\ref{cpll}) with  $m_L$
replaced by $m^*_{eff}$
This implies that the conditions relevant to discuss the
CP properties of the leptonic sector at low energies are
similar to those envolving $m_L$ and $h_l$ in 
Ref.\cite{Branco:1986gr} and can be translated into,
for instance:
\begin{eqnarray}
{\rm tr} \left[ (m_{eff} \;  m^*_{eff})^a,\; {h_l}^b \right]^q & = & 0
\label{prim} \\
{\rm Im \ tr }\left[ (h_l)^c \; (m_{eff} \; m^*_{eff})^d \;
(m_{eff}\; h^*_l \; m^*_{eff})^e \; (m_{eff} \; m^*_{eff})^f  \right] 
&= & 0 \label{seg} \\
{\rm Im \ det }\left[ ( m^*_{eff} \; h_l \; m_{eff}) \; + \; 
r (h^*_l \;  m^*_{eff} \; m_{eff} )  \right] &= & 0 \label{ter}
\end{eqnarray}
a, b, ..., f are integers, q is odd and r is an arbitrary real number.
These relations are necessary conditions for CP invariance. The
non-vanishing of any of these WB invariants implies CP violation.
However, these relations may not be sufficient to guarantee CP
invariance. In fact, there are cases where some of them vanish
automatically and yet CP may be violated. 

It is well known that the minimal structure that can lead to
CP violation in the leptonic sector is two generations of
left-handed Majorana neutrinos  requiring that 
their masses be non degenerate and that none of them
vanishes . In this case, it was proved \cite{Branco:1986gr} that
the condition
\begin{equation}
{\rm Im \ tr } \; Q = 0 \label{trq}
\end{equation}
with $Q = h_l  m_{eff}   m^*_{eff}  m_{eff} h^*_l  m^*_{eff}$
is a necessary and sufficient condition for CP invariance.

In the realistic case of three generations of light neutrinos
there are three independent CP violating phases relevant at 
low energies. In the physical basis they appear in the $V_{PMNS}$ 
matrix - one of them is a Dirac type phase analogous to
the one appearing in the Cabibbo, Kobayashi and Maskawa matrix,
$V_{CKM}$, 
of the quark sector and the two additional ones can be factored
out of $V_{PMNS}$ but cannot be rephased away due to the
Majorana character of the neutrinos.
Selecting from the necessary conditions a subset of 
restrictions which are also sufficient for CP invariance
is in general not trivial. For three generations it
was shown that the following four conditions are sufficient
\cite{Branco:1986gr} to guarantee CP invariance:
\begin{eqnarray}
{\rm Im \ tr } \left[ h_l \; (m_{eff} \; m^*_{eff}) \;
( m_{eff} \; h^*_l \; m^*_{eff})\right] & = & 0 \label{41} \\
{\rm Im \ tr } \left[ h_l \; (m_{eff} \; m^*_{eff})^2 \;   
( m_{eff} \; h^*_l \; m^*_{eff}) \right] & = & 0 \label{42} \\
{\rm Im \ tr } \left[ h_l \; (m_{eff} \; m^*_{eff})^2 \   
( m_{eff} \; h^*_l \; m^*_{eff}) \; (m_{eff} \; m^*_{eff})\right]
 & = & 0 \label{43} \\
{\rm Im \ det } \left[ ( m^*_{eff} \; h_l \; m_{eff}) 
+ (h^*_l \;  m^*_{eff} \; m_{eff} )\right]   & = & 0 \label{44} 
\end{eqnarray}
provided that neutrino masses are nonzero and nondegenerate.
It can be easily seen that these conditions are trivially
satisfied in the case of complete degeneracy 
$(m_1  = m_2 = m_3 ) $. Yet there may still be CP violation
in this case, as will be discussed in section 4.

Leptonic CP violation at low energies can be detected through
neutrino oscillations which are sensitive to the Dirac-type phase, 
but insensitive to the Majorana-type phases in  $V_{PMNS}$. 
In any given model, the strength of Dirac-type CP violation 
can be obtained from the
following low energy WB invariant:
\begin{equation}
Tr[h_{eff}, h_l]^3=6i \Delta_{21} \Delta_{32} \Delta_{31}
{\rm Im} \{ (h_{eff})_{12}(h_{eff})_{23}(h_{eff})_{31} \} \label{trc}
\end{equation}
where $h_{eff}=m_{eff}{m_{eff}}^{\dagger} $ and
$\Delta_{21}=({m_{\mu}}^2-{m_e}^2)$ with analogous expressions for
$\Delta_{31}$, $\Delta_{32}$. This invariant is, of course a
special case
of Eq. (\ref{prim}). For three left-handed neutrinos there is a Dirac-type 
CP violation if and only if this invariant does not vanish.
This quantity can be computed in any WB and can also be 
fully expressed in terms of physical observables since:
\begin{equation}
{\rm Im} \{ (h_{eff})_{12}(h_{eff})_{23}(h_{eff})_{31} \} =
- \Delta m_{21}^2 \Delta m_{31}^2  \Delta m_{32}^2 {\cal J}_{CP}
\label{fjcp} 
\end{equation}
where the $\Delta m_{ij}^2$'s are the usual light neutrino
mass squared differences and ${\cal J}_{CP}$ is the imaginary part 
of an invariant quartet of the leptonic mixing matrix $U_{\nu}$,
appearing in the difference of the CP-conjugated neutrino oscillation
probabilities, such as
$P(\nu_e\rightarrow\nu_\mu)-P(\bar{\nu}_e\rightarrow
\bar{\nu}_\mu)$. It is  given by:
\begin{equation}
{\cal J}_{CP} \equiv {\rm Im}\left[\,(U_\nu)_{11} (U_\nu)_{22}
(U_\nu)_{12}^\ast (U_\nu)_{21}^\ast\,\right] 
= \frac{1}{8} \sin(2\,\theta_{12}) \sin(2\,\theta_{13}) \sin(2\,\theta_{23})
\cos(\theta_{13})\sin \delta\,, \label{Jgen1}
\end{equation}
where the $\theta_{ij}$ , $\delta $
are the mixing angles and the Dirac-type phase appearing in the 
standard parametrization adopted in \cite{Eidelman:2004wy}.
The most salient feature of leptonic mixing is the fact that 
two of the mixing angles $(\theta_{12}, \theta_{23})$ are large,
with only $\theta_{13}$ being small. This opens the possibility 
of detecting leptonic CP violation through neutrino oscillations,
which requires ${\cal J}_{CP}$ to be of order $10^{-2}$, 
a value that can be achieved, provided $\theta_{13}$ 
is not extremely small
(at present one only has an experimental bound 
$\theta_{13} <  0.26 $).
A similar invariant condition is useful in the quark
sector \cite{Bernabeu:1986fc} where the corresponding
${\cal J}_{CP}$ is of the order $10^{-5}$. 
The search for CP violation
in the leptonic sector at low energies is 
at present one of the major experimental challenges in neutrino physics. 
Experiments with superbeams  and neutrino beams from
muon storage rings (neutrino factories)  have the potential \cite{CPobs} to
measure directly the Dirac phase $\delta$ through CP and $T$ asymmetries
or indirectly through oscillation probabilities which are
themselves  CP  conserving but also depend on $\delta$. An alternative method
\cite{farzan} is to measure the area of
unitarity triangles defined for the leptonic sector \cite{aguilar}.

\subsection{WB Invariants relevant for Leptogenesis}
One of the most plausible
scenarios for the generation of the baryon asymmetry 
of the Universe (BAU) is the leptogenesis
mechanism \cite{Fukugita:1986hr}
where a CP asymmetry generated through the
out-of-equilibrium L-violating decays of the heavy  
Majorana neutrinos leads to a lepton asymmetry which
is subsequently transformed into a baryon asymmetry
by (B+L)-violating sphaleron processes \cite{Kuzmin:1985mm}.

In this section, we consider thermal leptogenesis in the minimal
seesaw scenario. In what follows the notation will be
simplified into $m$ and $M$ for $m_D$ and $M_R$.
The lepton number asymmetry, $\varepsilon _{N_{j}}$, 
arising from the decay of the $j$th heavy Majorana
neutrino is defined in terms of the family number asymmetry
$\Delta {A^j}_i={N^j}_i-{{\overline{N}}^j}_i$ by:
\begin{equation}
\varepsilon _{N_{j}} 
= \frac{\sum_i \Delta {A^j}_i}{\sum_i \left({N^j}_i +
\overline{N^j}_i \right)}
\label{jad}
\end{equation}
the sum in $i$ runs over the three flavours
$i$ = e $\mu$ $\tau$.
The evaluation of $\varepsilon _{N_{j}}$, involves
the computation of the interference between the tree level
diagram and one loop diagrams for the decay of 
the heavy Majorana neutrino $N^j$
into charged leptons $l_i^\pm$ ($i$ = e, $\mu$ , $\tau$) 
which leads to
\cite{sym} :
\begin{eqnarray}
\varepsilon _{N_{j}}
&=& \frac{g^2}{{M_W}^2} \sum_{k \ne j} \left[
{\rm Im} \left((m^\dagger m)_{jk} (m^\dagger m)_{jk} \right)
\frac{1}{16 \pi} \left(I(x_k)+ \frac{\sqrt{x_k}}{1-x_k} \right)
\right]
\frac{1}{(m^\dagger m)_{jj}}   \nonumber \\
&=& \frac{g^2}{{M_W}^2} \sum_{k \ne j} \left[ (M_k)^2
{\rm Im} \left((R^\dagger R)_{jk} (R^\dagger R)_{jk} \right)
\frac{1}{16 \pi} \left(I(x_k)+ \frac{\sqrt{x_k}}{1-x_k} \right)
\right]
\frac{1}{(R^\dagger R)_{jj}} \nonumber \\
\label{rmy}
\end{eqnarray}
where $M_k$ denote the heavy neutrino masses,
the variable $x_k$
is defined as  $x_k=\frac{{M_k}^2}{{M_j}^2}$ and 
$ I(x_k)=\sqrt{x_k} \left(1+(1+x_k) \log(\frac{x_k}{1+x_k}) \right)$.
From Eq. (\ref{rmy})
it can be seen that the lepton-number
asymmetry is only sensitive to the CP-violating phases
appearing in $m^\dagger m$ in the WB, where $M_R \equiv M $
is diagonal (notice that this combination is insensitive
to rotations of the left-hand neutrinos).
Making use of the parametrization given by Eq. (\ref{upy})
for $m_D \equiv m $ 
it becomes clear that leptogenesis is only sensitive to
the phases $\beta_1$, $\beta_2$ and $\sigma$.
The second equality of Eq. (\ref{rmy})
is established  with the help of Eq. (\ref{app}).

Weak basis invariant conditions relevant for leptogenesis
must be sensitive to these three phases, clearly meaning
that they must be expressed in terms of  $h = m^\dagger m$.
From condition  Eq. (\ref{cpm}) we obtain 
\begin{equation}
W^{\prime \dagger} h W^{\prime} = h^* 
\end{equation}
Only the matrix $M$ is also sensitive to the $W^{\prime}$ rotation.
From condition Eq. (\ref{cpM}) we derive
\begin{equation}
W^{\prime \dagger} H W^{\prime} = H^* 
\end{equation}
where  $H = M^\dagger M$. From these two new conditions, together with 
Eq. (\ref{cpM}) it can be readily derived that CP invariance
requires \cite{Branco:2001pq}:
\begin{eqnarray}
I_1 \equiv {\rm Im Tr}[h H M^* h^* M]=0 \label{i1} \\
I_2 \equiv {\rm Im Tr}[h H^2 M^* h^* M] = 0 \label {i2l}\\
I_3 \equiv {\rm Im Tr}[h H^2 M^* h^* M H] = 0  \label{i3l}
\end{eqnarray}
as well as many other expressions of the same type. These 
conditions can be computed in any WB and are necessary and sufficient
to guarantee that CP is conserved at high energies.
This was shown by going to the WB where $M$ is real and diagonal.
In this basis the $I_i$'s are then of the form:
\begin{eqnarray}
I_1 &=& M_1 M_2 ({M_2}^2-{M_1}^2) {\rm Im}({h_{12}}^2)+M_1 M_3
({M_3}^2-{M_1}^2){\rm Im}( {h_{13}}^2) + \nonumber \\
&+& M_2 M_3 ({M_3}^2-{M_2}^2) {\rm Im}( {h_{23}}^2) \\
I_2 &=& M_1 M_2 ({M_2}^4-{M_1}^4) {\rm Im}({h_{12}}^2)+M_1 M_3
({M_3}^4-{M_1}^4){\rm Im}( {h_{13}}^2) + \nonumber \\
&+& M_2 M_3 ({M_3}^4-{M_2}^4) {\rm Im}( {h_{23}}^2) \\
I_3 &=&{M_1}^3 {M_2}^3 ({M_2}^2-{M_1}^2) {\rm Im}({h_{12}}^2)
+{M_1}^3 {M_3}^3 ({M_3}^2-{M_1}^2) {\rm Im}( {h_{13}}^2) + \nonumber \\
&+& {M_2}^3 {M_3}^3 ({M_3}^2-{M_2}^2) {\rm Im}( {h_{23}}^2)=0
\end{eqnarray}
These are a set of linear equations in terms of the variables 
${\rm Im}({h_{ij}}^2) = {\rm Im} 
 \left((m^\dagger m)_{ij} \right. $ 
$ \left.  (m^\dagger m)_{ij} \right)$ appearing
in Eq. (\ref{rmy}). The determinant of the coefficients of this
set of equations is:
\begin{equation}
Det={M_1}^2 {M_2}^2 {M_3}^2 
{\Delta^2}_{21}{\Delta^2}_{31}{\Delta^2}_{32}
\label{dmd}
\end{equation}
where ${\Delta}_{ij}=({M_i}^2-{M_j}^2)$. Non vanishing of the
determinant implies that all imaginary parts of $({h_{ij}})^2$
should vanish, in order for Eqs (\ref{i1}-\ref{i3l}) to hold.
Conversely, the non-vanishing of any of the $I_i$ implies
CP violation at high energies, relevant for leptogenesis.

\section{The Case of degenerate Neutrinos}
Since neutrino oscillations measure neutrino mass differences
and not the absolute mass scale, both hierarchical neutrino masses
and quasi-degenerate neutrino masses are allowed, by present 
experimental data.  
In the case of Dirac neutrinos, the limit of exact mass degeneracy
is trivial, since there is no mixing or CP violation in that limit.
The situation is entirely different for Majorana neutrinos, 
since in that case one can have both mixing and CP violation
even in the limit of exact degeneracy. 
The proof is simple \cite{Branco:1986gr} and follows from 
Eq. (\ref{ueff}) together with 
\begin{equation}
U^{\prime \dagger} h_l U^{\prime} = h^*_l   \label{abv} 
\end{equation}
which is readily obtained from Eq. (\ref{cpml}). 
Let us consider the low energy limit, where only 
left-handed neutrinos are relevant and assume that there are
three left-handed Majorana neutrinos with exact 
degenerate masses. Without loss of generality,
one can choose to work in a WB where the effective left-handed
neutrino mass matrix 
is diagonal, real. Since we are assuming the exact degeneracy limit,
the mass matrix is just proportional to the unit matrix.
We have seen that invariance under CP requires Eq. (\ref{ueff})
to be satisfied by some unitary matrix  $U^{\prime}$.
In the case of degeneracy and in the WB we have chosen, 
Eq. (\ref{ueff}) is satisfied provided that $ U^{\prime} = iO$ 
(with $O$ an orthogonal matrix). In addition we still have
the freedom to make a change of WB such that $m_{eff}$ is 
unchanged and Re$h_l$ becomes diagonal. In this basis 
Eq. (\ref{abv}) can be split into:
\begin{eqnarray}
O^T ({\rm Re } h_l ) O = {\rm Re } h_l \label{reh} \\
O^T ({\rm Im } h_l ) O = - {\rm Im } h_l \label{imh}
\end{eqnarray}
From Eq. (\ref{reh}) and assuming Re$h_l$ to be non degenerate
the matrix $O$ is constrained to be of the form 
$O = {\rm diag} ( \epsilon_1, .....,  \epsilon_n )$
with $ \epsilon_i = \pm 1 $. This in turn implies from  
Eq.~(\ref{imh}) that, in the general case of non vanishing 
$({\rm Im}h_l)_{ij}$, the  $ \epsilon_i$ have to obey 
the conditions:
\begin{equation}
\epsilon_i \cdot \epsilon_j = -1 \quad i \neq j    \label{eps} 
\end{equation}
Clearly these conditions cannot be simultaneously satisfied 
for more than two generations.

In the general case of three light neutrinos $V_{PMNS}$ can be
parametrized by three angles and three phases. In the limit of
exact degeneracy, in general mixing cannot be rotated away and 
$V_{PMNS}$  is parametrized by two angles and one CP violating
phase. We shall denote the corresponding leptonic mixing
matrix by $U_0$.
It has been shown \cite{Branco:1998bw}
that in general this matrix cannot be rotated away.
Only in the case where the theory is CP invariant and the three
degenerate neutrinos have the same CP parity can $U_0$ be rotated
away.

In the WB where the charged lepton mass matrix is diagonal,
real and positive the neutrino mass matrix is diagonalized
by the transformation
\begin{equation}
U^\dagger_0\cdot m_{eff} \cdot U^*_0\ =\ \mu   \cdot  {1\>\!\!\!{\rm I}}
\label{zzz} 
\end{equation}
where $\mu $ is the common neutrino mass. Let us define
the dimensionless matrix $Z_0=m_{eff} / \mu $. From Eq. (\ref{zzz})
we obtain:
\begin{equation}
Z_0\ =U_0 \cdot U_0^T \label{z0z0}
\end{equation}
which is unitary and symmetric. The matrix $Z_0$ can be
written without loss of generality as:
\begin{equation}
Z_0\ =\ \left(
\begin{array}{ccc}
1 & 0 & 0 \\
0 & c_\phi & s_\phi \\
0 & s_\phi & -c_\phi
\end{array}
\right) \cdot \left(
\begin{array}{ccc}
c_\theta & s_\theta & 0 \\
s_\theta & z_{22} & z_{23} \\
0 & z_{23} & z_{33}
\end{array}
\right) \cdot \left(
\begin{array}{ccc}
1 & 0 & 0 \\
0 & c_\phi & s_\phi \\
0 & s_\phi & -c_\phi
\end{array}
\right) 
\end{equation}
Unitarity of $Z_0$  implies that either $s_\theta $ or $z_{23}$
must vanish. The case $s_\theta =0$
automatically leads to CP invariance. Assuming $s_\theta \neq 0$ the
most general form for the symmetric unitary matrix $Z_0$ 
is then given by:
\begin{equation}
Z_0\ =\ \left(
\begin{array}{ccc}
1 & 0 & 0 \\
0 & c_\phi & s_\phi \\
0 & s_\phi & -c_\phi
\end{array}
\right) \cdot \left(
\begin{array}{ccc}
c_\theta & s_\theta & 0 \\
s_\theta & -c_\theta & 0 \\
0 & 0 & e^{i\alpha }
\end{array}
\right) \cdot \left(
\begin{array}{ccc}
1 & 0 & 0 \\
0 & c_\phi & s_\phi \\
0 & s_\phi & -c_\phi
\end{array}
\right) \label{neq}
\end{equation}
This choice of $Z_0$ does not include the
trivial case where CP is a good symmetry and all neutrinos have 
the same CP parity. In fact, in the CP conserving case 
where $e^{i\alpha } = \pm 1$ one has  
 ${\rm {Tr}}(Z_0) = - \det (Z_0) = \pm 1$ corresponding
to the eigenvalues $(1,-1,1)$ and $(1,-1,-1)$ and permutations.
It is well known \cite{Wolfenstein:1981rk} that different relative signs 
correspond to different CP parities. From Eqs. (\ref{z0z0}) and
(\ref{neq}) we conclude that the mixing matrix $U_0$
must be of the form:
\begin{equation}
U_0 \ =\ \left(
\begin{array}{ccc}
1 & 0 & 0 \\
0 & c_\phi & s_\phi \\
0 & s_\phi & -c_\phi
\end{array}
\right) \cdot \left(
\begin{array}{ccc}
\cos (\frac \theta 2) & \sin (\frac \theta 2) & 0 \\
\sin (\frac \theta 2) & -\cos (\frac \theta 2) & 0 \\
0 & 0 & e^{i\alpha /2}
\end{array} 
\right) \cdot \left(  
\begin{array}{ccc}  
1 & 0 & 0 \\
0 & i & 0 \\
0 & 0 & 1
\end{array}
\right) \label{uuuu}
\end{equation}
up to an arbitrary orthogonal transformation 
$U_0 \rightarrow U_0 \cdot O$ . Notice that $U_0 $ cannot 
be rotated away due to the fact that it is not an orthogonal
matrix, even in the CP conserving case. The matrix  $U_0 $
is parametrized by two angles
$\theta $, $\phi $ and one phase $\alpha $. In the limit of 
exact degeneracy a necessary and sufficient condition 
\cite{Branco:1998bw} for CP invariance is:
\begin{equation}
G\equiv \ {\rm {Tr}}\left[ \ (m^*_{eff}\cdot h_l\cdot m_{eff}\ ,\
h^*_l\right] ^3\ =\ 0
\end{equation}
In the WB where $h_l$ is diagonal 
i.e., $h_l=$diag$\ (m_e^2,\ m_\mu ^2,\ m_\tau ^2)$
it can be written as:
\begin{equation}
G=6i\ \Delta _m~{\rm {Im}}[(Z_0)_{11}^*(Z_0)_{22}^*
(Z_0)_{12}(Z_0)_{21}]=-\frac{3i}2\  
\Delta _m~\cos (\theta )\sin ^2(\theta )\ \sin ^2(2\phi )\ \sin (\alpha )
\label{gjg}
\end{equation}
where $\Delta _m=$\ $\mu ^6\ (m_\tau ^2-m_\mu ^2\ )^2(m_\tau ^2-m_e^2\
)^2(m_\mu ^2-m_e^2\ )^2$ is a multiplicative factor which contains the
different masses of the charged leptons and the common neutrino mass $\mu $.
One may wonder whether Eq. (\ref{uuuu}) would be a realistic 
mixing matrix for the case of three non degenerate
neutrinos. It has been shown \cite{Branco:1998bw}
that this is indeed the case. 
In fact this matrix corresponds to $\sin \theta_{13} =0$
(of the standard parametrization), solar neutrino data 
only  constrains  the angle $\theta$ 
whilst atmospheric neutrino data only constrains $\phi $. 
Neutrinoless double beta decay depends on $\theta$ and light
neutrino masses. The angle $\alpha$ can be factored out in 
$U_0$ and is thus a Majorana-type phase. 

Heavy Majorana
neutrinos may play a crucial r\^ ole in the generation of the
baryon asymmetry of the Universe. If these particles are indeed
responsible for BAU they must obey certain constraints (such as a
lower limit in their mass). It is common to assume
heavy neutrino masses to be hierarchical in the study 
of thermal leptogenesis since this corresponds to
the simplest scenario, which is sometimes called minimal
leptogenesis. Presently there are no direct
experimental constraints on
heavy neutrino masses, and the possibility of
quasi-degenerate heavy Majorana neutrinos remains open.

\section{On the relation between low energy CP violation 
and CP violation required for leptogenesis}

\subsection{Brief summary of low energy data}
There has been great experimental progress in the determination
of leptonic masses and mixing in the last few years. The 
evidence for solar and atmospheric neutrino oscillations is
now solid. The pattern of leptonic mixing ($V_{PMNS}$) is very
different from that of the quark sector ($V_{CKM}$), since only one 
of the leptonic mixing angles, $\theta_{13}$, is small.
The latest great progress reported is in the
measurement of the square mass difference relevant 
for solar oscillations, $\Delta m^2_{21} $, and is due to recent
KamLAND results \cite{kamland}. 
KamLAND is a terrestial long baseline experiment which
has great sensitivity to $\Delta m^2_{21} $, but it does not constrain 
$\theta_{12}$ much better than the current set of solar 
experiments. The combined result including
those of SNO \cite{sno} and previous solar
experiments \cite{nuexp} 
is for the 1$\sigma$ range \cite{parametros}:
\begin{eqnarray}
\Delta m^2_{21} & = & 8.2 ^{+0.3}_{-0.3} \times 10^{-5}\  {\rm eV}^2 \\
\tan ^2 \theta_{12} & = & 0.39 ^{+0.05}_{-0.04} 
\end{eqnarray}
and corresponds to the large mixing angle solution (LMA) of 
the Mikheev, Smirnov and Wolfenstein
(MSW) effect \cite{msw} with the upper island excluded.
On the other hand, atmospheric neutrino results from
Superkamiokande \cite{atm} and recent important progress by 
K2K \cite{k2k}, which is also a terrestrial long baseline experiment,
are consistent with, for the 1$\sigma$ range \cite{parametros}:
\begin{eqnarray}
\Delta m^2_{32} & = & 2.2 ^{+0.6}_{-0.4} \times 10^{-3}\  {\rm eV}^2 \\
\tan ^2 \theta_{23} & = & 1.0 ^{+0.35}_{-0.26} 
\end{eqnarray} 
The present bounds for  $\sin ^2  \theta_{13} $ from the 
CHOOZ experiment \cite{chooz} have been somewhat relaxed 
since they depend on $ \Delta m^2_{31} $ and this value 
went down. Assuming  the range for $ \Delta m^2_{32} $
from SuperKamiokande and K2K, 
the 3$\sigma$ bound \cite{parametros} lies in
$\sin ^2  \theta_{13} < 0.05-0.07 $. A higher value for the angle
$ \theta_{13} $ is good news for the prospectives of detection 
of low energy leptonic CP violation, mediated through a Dirac-type
phase, whose strength is given by ${\cal J}_{CP}$ defined in 
section 3. 
Direct kinematic limits on neutrino masses \cite{dir} 
from Mainz and Troitsk and
neutrinoless double beta decay experiments \cite{nonu}
when combined with the given square mass differences 
exclude light neutrino masses higher than order 1~eV. 
Non-vanishing light neutrino masses also have an 
important impact in
cosmology. Recent data from  the Wilkinson Microwave
Anisotropy Probe, WMAP \cite{wmap1}, \cite{wmap2}, 
together with other data, put an upper bound on the sum of light
neutrino masses of 0.7 eV. 

In the context of the seesaw mechanism the smallness of
light neutrino masses is related to the existence of
heavy neutrinos. These heavy neutrinos may in turn play
an important cosmological r\^ ole via the generation of
BAU through leptogenesis. Since leptogenesis requires
CP violation at high energies one may ask whether there
is a connexion between CP violation at low energies and
CP violation at high energies. This question will be addressed in the
next subsection.

\subsection{On the need for a lepton flavour symmetry}
The expression for the lepton-number
asymmetry resulting from the decay of heavy Majorana
neutrinos is given by Eq. (\ref{rmy}). Yet leptogenesis
is a complicated thermodynamical non-equilibrium
process and depends on additional parameters. The simplest 
scenario corresponds to heavy hierarchical neutrinos 
where $M_1$ is much smaller than $M_2$ and $M_3$. The case
of almost degeneracy of heavy neutrinos has been considered
by several authors \cite{resonance} and corresponds to a 
resonant enhancement of $\varepsilon_{N_j}$. In the 
hierarchical case the baryon asymmetry only depends on four
parameters \cite{Buchmuller}: the mass $M_{1}$ of the lightest heavy
neutrino, together with the corresponding 
CP asymmetry $\varepsilon _{N_{1}}$ in
their decays, as well as the effective neutrino mass $\widetilde{m_{1}}$  
defined as
\begin{equation}
\widetilde{m_{1}}=(m^{\dagger }m)_{11}/M_{1}  \label{mtil}
\end{equation}
in the weak basis where $M$ is diagonal, real and positive and, finally, the
sum of all light neutrino masses squared, 
${\bar{m}}^{2}=m_{1}^{2}+m_{2}^{2}+m_{3}^{2}$. It has been shown that this
sum controls an important class of washout processes.
Successful leptogenesis would require $\varepsilon _{N_{1}} $ 
of order $10^{-8}$,
if washout processes could be neglected, in order to reproduce
the observed ratio of baryons to photons 
\cite{wmap1}:
\begin{equation}
\frac{n_{B}}{n_{\gamma}}= (6.1 ^{+0.3}_{-0.2}) \times 10^{-10}.
\end{equation}  
Leptogenesis is a non-equilibrium process
that takes place at temperatures $T\sim M_{1}$.
This imposes an upper bound on the effective neutrino mass 
$\widetilde{m_{1}}$ given by the ``equilibrium neutrino mass''
\cite{kfb}:
\begin{equation} 
m_{*}=\frac{16\pi ^{5/2}}{3\sqrt{5}}g_{*}^{1/2}\frac{v^{2}}{M_{Pl}}\simeq
10^{-3}\ \mbox{eV}\; ,  \label{enm}
\end{equation}  
where $M_{Pl}$ is the Planck mass ($M_{Pl}=1.2\times 10^{19}$ GeV),
$v=\langle \phi ^{0}\rangle /\sqrt{2}\simeq 174\,$GeV is the weak scale 
and $g_{*}$ is the effective number of relativistic degrees of
freedom in the plasma and equals 106.75 in the SM case.
Yet, it has been shown \cite{bdbp} that successful leptogenesis
is possible for $\widetilde{m_{1}} < m_{*}$ as well as
$\widetilde{m_{1}} > m_{*}$, in the range from 
$\sqrt {\Delta m^2_{12}}$ to  $\sqrt {\Delta m^2_{23}}$. 
The square root of the sum
of all neutrino masses squared ${\bar{m}}$ is constrained, in the
case of normal hierarchy, to be below 0.20 eV \cite{bdbp}, 
which corresponds to an upper bound on light neutrino masses
very close to 0.10 eV. This result is sensitive to radiative
corrections which depend on top and Higgs masses as well as on
the treatment of thermal corrections.  
In \cite{gnrrs} a slightly higher value of 0.15 eV is found.
This bound can be relaxed for instance in various scenarios 
including models with quasi degenerate heavy neutrinos 
\cite{resonance},  non thermal leptogenesis
scenarios \cite{nonth}, or also theories with Higgs triplets 
\cite{triple} leading to non-minimal seesaw mechanism.
In the limit $M_{1}\ll M_{2},M_{3}$,  $\varepsilon _{N_{1}} $ can
be simplified  into:
\begin{equation}
\varepsilon _{N_{1}}\simeq -\frac{3}{16\,\pi v^{2}}\,\left( I_{12}\,
\frac{M}{M_{2}}+I_{13}\,\frac{M_{1}}{M_{3}}\right) \,,  \label{lepto3}
\end{equation}
where
\begin{equation}
I_{1i}\equiv \frac{\mathrm{Im}\left[ (m^{\dagger }m)_{1i}^{2}\right] }
{(m^{\dagger }\,m)_{11}}\ .  \label{lepto4}
\end{equation}
and a lower bound on the lightest heavy neutrino mass $M_{1}$
is derived. Depending on the cosmological scenario, the range for
minimal  $M_{1}$ varies from order $10^7$ Gev to $10^9$ Gev 
\cite{Buchmuller} \cite{gnrrs}.

Viability of leptogenesis is thus closely related to low
energy parameters, in particular the light neutrino masses.
This raises the question of whether the same is true for
CP violation at both low and high energies. Part of the
answer to this question  \cite{Buchmuller:1996pa}
is given here in section 3.2 where it was shown 
that leptogenesis only depends on the phases
$\beta_1$, $\beta_2$ and $\sigma $ whilst the phases in $V_{PMNS}$ 
depend on all six phases \cite{Branco:2001pq}. 
The question remais of whether
a CP conserving low energy theory (no Dirac-type and no 
Majorana-type phases) would still allow for high energy CP violation.
The answer is yes \cite{Rebelo:2002wj}, 
since the matrix $m$ can be parametrized
in such a way that $V_{PMNS}$ cancels out in 
the product $m^{\dagger }m$ and all the additional phases remaining 
in this product cancel out in $m_{eff}$.
As a result, any connection between CP violation at low and at
high energies is model dependent. 
More specifically, in order to establish the above connection,
one has to restrict the number of free parameters in the lepton 
flavour sector. An elegant way of obtaining such restrictions is
through the introduction of a lepton-flavour symmetry. There
is another motivation for restricting the number of free parameters
in the lepton flavour sector. This has to do with the fact that,
contrary to what happens in the quark sector, without lepton 
flavour restrictions, it is not possible to fully reconstruct
the low energy neutrino mass matrix from low energy data
obtainable through feasible experiments \cite{feasi}.

Several authors have studied
the connection between CP violation at low and at high energies  
in various interesting scenarios  \cite{CPconnexion}. 
An important motivation for such studies is the attempt 
to show whether or not the baryon asymmetry of the Universe 
was generated through leptogenesis.

\subsection{Towards a minimal Scenario}
A particular minimal scenario allowing to establish a link between BAU
generated through leptogenesis and CP violation at low energies
was considered in Ref. \cite{Branco:2002xf}. The starting point 
was to write $m$, the Dirac type neutrino mass matrix,
as the product of a unitary times a lower triangular matrix 
in the weak basis where $M$ and $m_l$ are diagonal and real.
As pointed out before there is no lack of generality in choosing
this parametrization. The strategy was then to simplify this matrix
$m$ in order to obtain physical constraints. Starting from:
\begin{equation}
m_D = U\,Y_{\triangle}\,, \label{mDtri} 
\end{equation}
with $Y_{\triangle}$ of the form:
\begin{equation}
Y_{\triangle}= \left(\begin{array}{ccc}
y_{11} & 0 & 0 \\
y_{21}\,e^{i\,\phi_{21}} & y_{22} & 0 \\
y_{31}\,e^{i\,\phi_{31}} & y_{32}\,e^{i\,\phi_{32}} & y_{33}
\end{array}
\right)\,, \label{Ytri1}
\end{equation}
where $y_{ij}$ are real positive numbers, it follows that $U$
does not play any r\^ ole for leptogenesis since it cancels out in
the product $m^\dagger m$. It is clear that a necessary condition
for a direct link between leptogenesis and low energy
CP violation to exist is the requirement that the matrix $U$
contains no CP violating phases. The simplest possible choice,
corresponding to $U= {1\>\!\!\!{\rm I}} $, was made. 
Next, further simplifying restrictions were imposed on
$Y_{\triangle}$ in order to obtain minimal scenarios based on the
triangular decomposition. These correspond to special
zero textures together with assumptions on the
hierarchy of the different entries. Only two patterns with
one additional zero in $Y_{\triangle}$ where found to be
consistent with low energy physics (either with hierarchical 
heavy neutrinos or two-fold quasi degeneracy):
\begin{equation}
\left(\begin{array}{ccc}
 y_{11}   & 0       &0 \\
 y_{21}\,e^{i\,\phi_{21}}  & y_{22}       & 0 \\
0   & y_{32}\,e^{i\,\phi_{32}} & y_{33}
\end{array}\right) \, ,\qquad 
\left(\begin{array}{ccc}
 y_{11}   & 0       & 0 \\
 0  & y_{22}       & 0 \\
 y_{31}\,e^{i\,\phi_{31}}   & y_{32}\,e^{i\,\phi_{32}} & y_{33}
\end{array}\right)
\label{duas}
\end{equation}
In both cases there are two independent phases. A further 
simplification is to assume one of these phases to vanish. Special 
examples were built and it was shown that it is possible
to obtain viable leptogenesis in this class of models
and at the same time obtain specific predictions for
low energy physics once the known experimental constraints are
imposed. In particular all the textures considered predicted the
existence of low energy CP violating effects in the range of
sensitivity of future long baseline experiments. It should 
be noted that strong hierarchies in the entries of masses matrices
could in principle be generated by the Froggatt-Nielsen 
mechanism \cite{Froggatt:1978nt}.

The question of whether the sign of the baryon asymmetry 
of the Universe can be related to CP violation in neutrino 
oscillation experiments was addressed by considering 
models with only two heavy neutrinos \cite{Frampton:2002qc}. 
In this case the Dirac mass
matrix has dimension $3 \times 2 $. The interesting examples 
correspond to textures of the form given above in Eq. (\ref{duas})
with the third column eliminated and corresponds to the
most economical extension of the SM leading to leptogenesis.
With the elimination of the third column one more phase 
in the third row can be rotated away, 
hence only one physical phase remains. 
In fact, there are fewer parameters in this case and these are 
strongly constrained by low energy physics thus leading to
a definite relative sign between Im $(m^\dagger m)^2_{12}$
and $\sin 2 \delta $ (with $\delta$ the Dirac type phase of
$V_{PMNS}$).

\section{Summary and Conclusions}
We have reviewed leptonic CP violation and neutrino mass models, 
with emphasis on the use of WB invariants to study CP violation 
at low and high energies, as well as on the possible connection 
between leptonic CP violation at low energies and CP violation 
required for the generation of the baryon asymmetry of the 
Universe through leptogenesis. We have identified the WB 
invariant which measures the strength of Dirac-type CP 
violation at low energies for three generations of light neutrinos
and have presented the simplest WB invariants which are sensitive
to CP violation required by leptogenesis. These WB invariants 
are specially relevant for the study of any given lepton-flavour
model, where Yukawa couplings are constrained by
lepton-flavour symmetries leading, for example, to
texture zeros in the leptonic mass matrices. The usefulness 
of the invariants stems from the fact that they can be
applied in any WB, without having to perform any cumbersome change 
of basis.

Most of our analysis was done in the framework of the
minimal seesaw mechanism, where there is a closer connection
between low energy data and leptogenesis. We have also 
considered some special cases such as the limit of exact 
degeneracy, illustrating the fact that for three Majorana neutrinos, 
both leptonic mixing and CP violation can exist even
in the limit where neutrinos are exactly degenerate.

In conclusion, neutrino physics provides an invaluable tool
to the study of the question of leptonic flavour and CP
violation at low energies, while at the same time having
profound implications to the physics of the early
universe, in particular to the generation of the baryon asymmetry 
of the Universe.

\end{document}